\documentclass[prd,nofootinbib]{revtex4}
\usepackage{graphicx, epsfig}
\usepackage{color}
\usepackage{mathrsfs}
\usepackage{amsmath}

\newcommand{\be}{\begin{equation}}
\newcommand{\ee}{\end{equation}}
\newcommand{\bd}{\begin{equation*}}
\newcommand{\ed}{\end{equation*}}
\newcommand{\bea}{\begin{eqnarray}}
\newcommand{\eea}{\end{eqnarray}}
\newcommand{\vx}{\vec{x}}
\newcommand{\vk}{\vec{k}}
\newcommand{\hs}{\hspace{2mm}}

\newcommand{\gapp}{\mathrel{\raise.3ex\hbox{$>$}\mkern-14mu
              \lower0.6ex\hbox{$\sim$}}}
\newcommand{\lapp}{\mathrel{\raise.3ex\hbox{$<$}\mkern-14mu
              \lower0.6ex\hbox{$\sim$}}}

\begin{document}

\title{Time dependent particle production and particle number in cosmological de Sitter space}
\author{Eric Greenwood}
\affiliation{New College of Florida, Sarasota, FL, 34243, U.S.A.}
\begin{abstract}
In this paper we consider the occupation number of induced quasi-particles which are produced 
during a time-dependent process using three-different methods: Instantaneous diagonalization, 
the usual Bogolyubov transformation between two different vacua (more precisely the instantaneous 
vacuum and the so-called adiabatic vacuum), and the Unruh-de Witt detector methods. Here we 
consider the Hamiltonian for a time-dependent Harmonic oscillator, where both the mass and 
frequency are taken to be time-dependent. From the Hamiltonian we derive the occupation number 
of the induced quasi-particles using the invariant operator method; in deriving the occupation number 
we also point out and make the connection between the Functional Schr\"odinger formalism, 
quantum kinetic equation, and Bogolyubov transformation between two different Fock space 
basis at equal times and explain the role in which the invariant operator method plays. 

As a concrete example, we consider particle production in the flat FRW chart of de Sitter spacetime. Here 
we show that the different methods lead to different results: The instantaneous diagonalization 
method leads to a power law distribution, while the usual Bogolyubov transformation and 
Unruh-de Witt detector methods both lead to thermal distributions (however the dimensionality 
of the results are not consistent with the dimensionality of the problem; the usual Bogolyubov 
transformation method implies that the dimensionality is $3D$ while the Unruh-de Witt 
detector method implies that the dimensionality is $7D/2$). It is shown that the source of the 
descrepency between the instantaneous diagonalization and usual Bogolyubov methods is 
the fact that there is no notion of well-defined particles in the out vacuum due to a divergent 
term. In the usual Bogolyubov method, this divergent term cancels leading to the thermal 
distribution, while in the instantaneous diagonalization method there is no such cancelation 
leading to the power law distribution. However, to obtain the thermal distribution in the usual Bogolyubov 
method, one must use the large mass limit. On physical grounds, one should expect that only the modes 
which have been allowed to sample the horizon would be thermal, thus in the large mass limit these modes 
are well within the horizon and, even though they do grow, they remain well within the horizon due to 
the mass. Thus one should not expect a thermal distribution since the modes won't have a chance to 
thermalize.
\end{abstract}

\maketitle

\section{Introduction}

The interest in quantum field theory in classical curved space-time, that is adding quantum fields to the 
classical gravitational background, has since particle production was first investigated in the cosmological 
context by L.~Parker in 1968 \cite{Parker,Parker2}, and black hole evaporation by S. W. Hawking in 1974 
\cite{Hawking:1974rv} persisted till today. Unlike the study of quantum field theory in Minkowski 
space-time, there are many difficulties and ambiguities associated with the study of quantum field theory in 
curved space-time. For example, it is well-known that the concept of a particle is not well defined in curved space-time, since there is no notion of a unique observer for any given space-time \cite{Birrell:1982ix}. 
Thus, the question of how defining of the vacuum state (or equivalently how many particles are present) 
becomes an observer dependent question. 

As an attempt 
to surmount this difficulty, many authors have proposed different methods for determining the particle 
content of a particular space-time, ranging from instantaneously diagonalizing the 
Hamiltonian \cite{Grib:1976pw}, particle detectors \cite{Unruh:1975gz}, and Bogolyubov 
transformation \cite{Bogolyubov:1980nc} to name a few. These calculations of particle production 
usually deal with comparing the particle number with respect to vacuum states defined in two different 
frames or in terms of the time-dependent, one-mode occupation number. Ideally, as long as one is consistent 
in defining the observer that one wishes to consider, each of these methods should give the same 
results. 

In a general time-dependent space-time, the investigation of the spectrum of induced quasi-particles 
reduces to the study of a harmonic oscillator with time-dependent parameters. Whether one is 
comparing the particle number with respect to two different vacuum states or solving for the time-dependent, 
one-mode occupation number, one must solve the time-dependent harmonic oscillator to determine the 
time-dependent mode amplitudes. One powerful technique for solving the time-dependent harmonic 
oscillator is to utilize the method of invariants \cite{Lewis:1968yx,Lewis:1968tm}, which  
allows one to quantize a harmonic oscillator with time dependent coefficients. 

The point of this paper is three-fold. First, to derive the semi-classical occupation number using three 
different methods. Second, to make clear the connection between the Functional Schr\"odinger formalism 
(FSF) \cite{Vachaspati:2006ki,Greenwood:2010mr,Greenwood:2010sx}, 
quantum kinetic \cite{Podolsky:2012wn}, that is the method for determining the time-dependent particle spectrum originally derived by Zeldovich and Starobinsky in 1971 and Mamayev {\it et.~al.}~in 1977 
\cite{Zeldovich:1971mw,Mamaev:1977pn} and the central role in which the invariant operator method 
plays in these formalisms. Third, to give a practical example and compare the results of the three different 
methods. 

As a practical example, we will consider a time-dependent space-time of cosmological interest, that of 
FRW. Here we will investigate what a particular observer, an observer in the flat 
FRW chart in de Sitter space-time, measures as the occupation number, hence the particle 
content, using each of these different methods. When comparing the three different methods we will see that 
all three of these methods lead to different results. The naive expectation is that one should acquire a thermal 
distribution, in accord with the celebrated work of Gibbons and Hawking \cite{Gibbons:1977mu}.

The paper is organized as follows: 
In Section \ref{sec:Particle} we derive the occupation number for a scalar field with a 
time-dependent ``mass" and frequency using three different methods: The instantaneous 
diagonalization of the associated Hamiltonian for the scalar field, the usual Bogolyubov 
transformation method between two different basis functions, corresponding two two different 
vacua and usually two different observers, and the Unruh-de Witt detector method.\footnote{In 
general there is no simple relationship between the occupation number and the particle 
number as measured by a detector, even in the freely falling case. To check, one should 
calculate $\langle T_{\mu\nu}\rangle$ and verify that there is also a thermal bath of particles 
in the out region \cite{Birrell:1982ix}. For example, as is well-known, when considering an 
uniformly accelerated observer in Minkowski space, the detector registers a thermal bath, however 
you finds that the expectation value of the stress energy tensor is in fact zero.} In this section we will make 
clear the connection between the invariant operator and the occupation number derived in the context 
of the FSF, as defined as the Gaussian overlap between the final and initial states, in the context of 
a time-dependent Bogolyubov transformation between the vacuum state of the harmonic oscillator and 
that of the invariant operator, and the quantum kinetic equations of Zeldovich and Starobinsky, as well 
as the role of the Ermokov equation to determine the quasi-particle spectrum. 
In Section \ref{sec:FRW} we derive the equations of motion for the mode coefficients of a scalar 
field in an arbitrary $4$-dimensional Friedman-Robertson-Walker space-time. In Section 
\ref{sec:dS prod} we specialize to the de Sitter space-time and derive the distribution for the 
occupation associated with the three different methods. In Section \ref{sec:Discussion} 
we conclude with some discussion and comments. 

\section{Particle Creation: Time-dependent Harmonic Oscillator}
\label{sec:Particle}

In this section we will consider particle creation, induced quasi-particle production for a general 
time-dependent Hamiltonian were both the ``mass" and the frequency are time-dependent. 
To determine the spectrum of the induced quasi-particles, we will derive the occupation 
number using three different methods: First, by instantaneously diagonalizing the Hamiltonian 
at any given moment of time. Second, by considering the usual Bogolyubov transformation 
method between two different basis functions for the same scalar field, which usually amounts 
to considering the same scalar field as seen by two different observers. Finally, by considering 
the response rate of a simple monopole detector, the Unruh-de Witt detector, which is following 
some trajectory within the space-time.

\subsection{Instantaneous Diagonalization}
\label{sec:Inst}

To consider the particle creation at any moment of time, we will consider the Hamiltonian 
of the induced quasi-particles, which is generically given by
\be
  H=\frac{1}{2M(t)}\Pi_k^2+\frac{1}{2}M(t)\omega^2(t)q_k^2
  \label{Ham}
\ee
where $\Pi_k$ is the conjugate momentum to the coordinate $q_k$, $M(t)$ is a time-dependent 
``mass" of the induced quasi-particle, and $\omega(t)$ is a time-dependent frequency. 
In determining the occupation number of the induced quasi-particles, we will work in the Heisenberg 
picture\footnote{We could equal well work in the Interaction picture, where the vacuum states are 
time-dependent as well. Here the creation operator defined in (\ref{Ai}) are related to the creation 
operators in the Interaction picture by
\bd
  \hat A_k(t)=\hat{\tilde A}_k(t)\exp\left[-i\int_{t_0}^t\omega(t')dt'\right].
\ed
}. Here we define new operators, which diagonalize the Hamiltonian at all moments of time
\be
  \hat A_k=\frac{1}{\sqrt{2M\omega}}\left(\Pi_k-iM\omega q_k\right).
  \label{Ai}
\ee
These operators satisfy the commutation relation 
$[\hat A_k,\hat A^{\dagger}_k]=\delta^{(3)}(k-k')$, as well as the Heisenberg equation
\be
  \frac{d\hat A_k}{dt}=-\frac{1}{2M\omega}\frac{d}{dt}(M\omega)\hat A^{\dagger}_k+i\left[H(\hat A_k,\hat A^{\dagger}_k),\hat A_k\right].
  \label{dAdt}
\ee
In (\ref{dAdt}) the first term describes the time-dependence due to the redefinition of the notion of 
particles for every moment $t$. Using these operators one can derive the kinetic equation,
see \cite{Schmidt:1998vi} for details, which is an integral-differential equation describing the particle 
production during the time-dependent process. The vacuum associated with these operators is defined 
by $\hat A_k|0\rangle_A=0$ for all $k$, or considering the wavefunctional 
$\varphi(q_k)=\langle q_k|0\rangle_A$, which leads to the normalized wavefunctional
\be
  \varphi(q_k)=\left(\frac{M\omega}{\pi}\right)^{1/4}e^{-M\omega q_k^2/2}
  \label{varphi}
\ee
which is just the harmonic oscillator ground state wavefunctional as expected from the structure 
of the Hamiltonian. (\ref{varphi}) corresponds to the usual positive frequency Minkowski vacuum 
state \cite{Kluson:2003sh}.

However, one can use the invariant operator method to study the time-dependence of the 
quantum system 
\cite{Lewis:1968yx,Lewis:1968tm}. Here, one defines a Hermitian invariant operator that satisfies 
\bd
  \frac{dI}{dt}=\frac{\partial I}{\partial t}-i[I,H]=0,
\ed
which has real, time-independent, eigenvalues. In our case the invariant can be decomposed in 
terms of linear invariants given as
\bd
  \hat c_k=\frac{1}{\sqrt{2}}\left[\rho \Pi_k-M\frac{d\rho}{dt}q_k-i\frac{q_k}{\rho}\right],
\ed
where $\rho$ is the real solution to the auxiliary equation
\be
  \frac{d^2\rho}{dt^2}+\sigma\frac{d\rho}{dt}+\omega^2(t)\rho=\frac{1}{M^2\rho^3},
  \label{rho EOM}
\ee
where 
\bd
  \sigma=\frac{d}{dt}\ln(M),
\ed
with initial conditions 
\be
  \rho(t=t_0)=\sqrt{\frac{1}{M_0\omega_0}},\hs \text{and} \hs \frac{d\rho}{dt}(t=t_0)=0
  \label{rho IC}
\ee 
where $M_0$ is the ``mass" at $t=t_0$ and $\omega_0$ is the frequency at $t=t_0$.
The quadratic, Hermitian, invariant operator is then given by 
$I(t)=\left(\hat c^{\dagger}_k \hat c_k+\frac{1}{2}\right)$
where the creation and annihilation operators satisfy 
$[\hat c_k,\hat c^{\dagger}_{k'}]=\delta^{(3)}(k-k')$.
The vacuum associated with these operators are defined by $\hat c_k|0\rangle_c=0$, or 
considering $\psi(q_k,t)=\langle q_k|0\rangle_c$, leads to the 
normalized wavefunctional\footnote{We recognize the wavefunctional (\ref{psi}) as the solution to 
the functional Schr\"odinger 
formalism as found in \cite{Greenwood:2010mr}}
\be
  \psi(q_k,t)=\left(\frac{1}{\pi \rho^2}\right)^{1/4}\exp\left[\frac{iM}{2}\left(\frac{1}{\rho}\frac{d\rho}{dt}+\frac{i}{M\rho^2}\right)q_k^2\right].
  \label{psi}
\ee
This vacuum is distinct from the harmonic oscillator vacuum in that the operators $\hat A_k$ 
and $\hat c_k$ are related through a Bogolyubov transformation \cite{Bertoni:1997qb}
\be
  \hat A_k=\mu_k(t)\hat c_k+\nu_k(t)\hat c^{\dagger}_k,
  \label{bog trans}
\ee
where 
\begin{align}
  \mu_k(t)=\frac{1}{2\sqrt{M\omega}}\left(\rho M\omega+\frac{1}{\rho}+iM\frac{d\rho}{dt}\right),\hs \text{and} \hs \nu_k(t)=-\frac{1}{2\sqrt{M\omega}}\left(\rho M\omega-\frac{1}{\rho}+iM\frac{d\rho}{dt}\right).
  \label{bog coeff}
\end{align}
We note that at the initial time $t=t_0$, $\mu_k(t=t_0)=1$ and $\nu_k(t=t_0)=0$ and the 
Bogolyubov transformation satisfies $|\mu_k(t)|^2-|\nu_k(t)|^2=1$ for all time $t$.

The transformation (\ref{bog trans}) is between two different Fock space basis at equal times, 
not between the same basis at different times. From 
(\ref{bog trans}) we see that at time $t=t_0$, $\hat A_k=\hat c_k$ and 
$\psi(q_k,t=t_0)=\varphi(q_k)$ 
so that the creation and annihilation operators for the two vacua are equivalent and the vacua are 
equivalent, thus there is no mixing at the initial time. 

The occupation number of particles created during the time of expansion then amounts to 
determining the number of $\hat A_k$ particles in the vacuum\footnote{Equivalently one can 
consider the number of $\hat c_k$ particles in the vacuum $|0\rangle_A$. The physical difference 
between the two descriptions is whether one is working in the ``in-in" picture 
versus the ``out-out" picture. Here, the vacuum $|0\rangle_A$ corresponds to the ``in" vacuum, 
while the vacuum $|0\rangle_c$ corresponds to the ``out" vacuum.} $|0\rangle_c$: 
\be
  N={}_c\langle0|\hat A^{\dagger}_k\hat A_k|0\rangle_c=|\nu_k(t)|^2.
  \label{N def}
\ee
It is then easy to show that the spectrum of instantaneous excitations from (\ref{N def}) is given 
by\footnote{The exact form of the spectrum was previously derived in the 
context of the functional Schr\"odinger formalism \cite{Greenwood:2010mr}.}
\be
  N(\omega_k,t)=\frac{M\omega\rho^2}{4}\left[\left(1-\frac{1}{M\omega\rho^2}\right)^2+\left(\frac{1}{\omega\rho}\frac{d\rho}{dt}\right)^2\right].
  \label{N}
\ee

Before we move on, we will comment on (\ref{N}) and the relationship between the functional 
Schr\"odinger formalism (FSF) and the Bogolyubov method considered here and elsewhere. 
Here we note that (\ref{N}) is exactly the spectrum for the occupation number found using (FSF), see 
\cite{Greenwood:2010mr} and references therein for applications to black hole 
physics. In FSF, the spectrum for the occupation number is derived by considering the Gaussian 
overlap between the initial vacuum wavefunctional, here the ``in" vacuum wavefunctional given in 
(\ref{varphi}), and the wavefunctional of the invariant operator at a later time, here the ``out" 
vacuum wavefunctional given in 
(\ref{psi}), at a given frequency $\bar\omega$, which is the frequency at the given time $t_f$. 
Thus the Gaussian overlap in the context of first 
quantization amounts to an equal time Bogolubov transformation between two different Fock state 
basis in the context of second quantization. 

As an alternative method for determining the number of induced quasi-particles one can use the fact 
that, as noted in Section \ref{sec:Inst} below (\ref{bog coeff}), the Bogolyubov coefficients satisfy 
the relation $|\mu_k|^2-|\nu_k|^2=1$, hence we can find the number of particles to 
\be
  N=\frac{1}{|\zeta_k|^2-1}
  \label{op N}
\ee
where
\be
  \zeta_k\equiv\frac{\mu_k}{\nu_k}=\frac{1+iM\rho\dot\rho+M\omega\rho^2}{1-iM\rho\dot\rho-M\omega\rho^2} 
  \label{zeta}
\ee
where $\dot\rho=d\rho/dt$ and we used (\ref{bog coeff}).\footnote{From (\ref{zeta}) we then have
\bd
  |\zeta_k|^2=\frac{M^2\rho^2\dot\rho^2+(M\omega\rho^2+1)^2}{M^2\rho^2\dot\rho^2+(m\omega\rho^2-1)^2}
\ed 
which could be used equally well.} Of course (\ref{op N}) leads to the same occupation as in (\ref{N}) when one 
simplifies this expression.

Furthermore, using (\ref{Ai}), (\ref{dAdt}) and their Hermitian conjugates, one can derive 
the three coupled first-order differential equations
\begin{align}
  \frac{dN}{dt}&=-\frac{1}{2M\omega}\frac{d}{dt}(M\omega)\left(\sigma^{\dagger}+\sigma\right),\nonumber\\
  \frac{d\sigma}{dt}&=-2i\omega\sigma-\frac{1}{2M\omega}\frac{d}{dt}(M\omega)(1+2N),\nonumber\\
  \frac{d\sigma^\dagger}{dt}&=2i\omega\sigma^\dagger-\frac{1}{2M\omega}\frac{d}{dt}(M\omega)(1+2N),
  \label{Nchi}
\end{align}
where we have defined the operator $N=\hat A_k^\dagger\hat A_k$ and an anomalous operator 
$\sigma=\hat A_k\hat A_k$. We can now define new variables $X=\sigma+\sigma^\dagger=2\Re(\sigma)$ 
and $Y=i(\sigma-\sigma^\dagger)=2i\Im(\sigma)$ and rewrite (\ref{Nchi}) as
\begin{align}
  \frac{dN}{dt}&=-\frac{1}{2M\omega}\frac{d}{dt}(M\omega)X,\nonumber\\
  \frac{dX}{dt}&=-2\omega Y-\frac{1}{M\omega}\frac{d}{dt}(M\omega)(1+2N),\nonumber\\
  \frac{dY}{dt}&=2\omega X,
  \label{NXY}
\end{align}
which are the same as the first-order differential equations originally considered by Zeldovich 
and Starobinsky, and Mamayev {\it et.~al.}~\cite{Zeldovich:1971mw,Mamaev:1977pn}. 
These operators can instead be defined as 
combinations of the Bogolyubov coefficients in (\ref{bog coeff}): Here
if one defines along with $N(\omega,t)$ in (\ref{N}) the combinations 
\bd
  X(\omega,t)={}_c\langle0|(\hat A_k\hat A_k+\hat A^\dagger_k\hat A^\dagger_k)|0\rangle_c=\mu_k(t)\nu_k(t)+\mu^*_k(t)\nu^*_k(t)=(2N+1)-M\omega\rho^2,
\ed 
and 
\bd
  Y(\omega,t)=i{}_c\langle0|(\hat A_k\hat A_k-\hat A^\dagger_k\hat A^\dagger_k)|0\rangle_c=i(\mu_k(t)\nu_k(t)-\mu^*_k(t)\nu^*_k(t))=M\frac{d\rho}{dt}\rho=\frac{1}{2}M\frac{d}{dt}\rho^2, 
\ed
then the Hamiltonian in (\ref{Ham}) is 
diagonalized if these combinations satisfy the first-order differential equations in (\ref{NXY}).
Initial conditions for (\ref{NXY}) are 
\be
  N(\omega_0,t=t_0)=X(\omega_0,t=t_0)=Y(\omega_0,t=t_0)=0.
  \label{NXY IC}
\ee
In fact, it is possible to eliminate $X$ and $Y$ from (\ref{NXY}) and write a single integro-differential 
equation for the operator $N$, which is the quantum kinetic equation describing the production of 
the instantaneous quasi-particles, see \cite{Schmidt:1998vi}. Numerically if one solves the system of coupled 
equations, one obtains the same spectrum for the occupation number of the induced quasi-particles 
obtained from FSF and the Bogolyubov transformation between the two different Fock space basis.

To find the 
spectrum of the occupation number, one must solve for the auxiliary equation for $\rho$.
It was originally reported in \cite{Greenwood:2010mr}, as well as elsewhere, that 
(\ref{rho EOM}) must be solved numerically due to the non-linear nature of the equation. 
However, it turns out that this not necessarily true; using the Ermakov equations 
\cite{Kevrekidis} the relationship between the time-dependent amplitude $q_k$ and 
$\rho$ is given by $q_k=\rho e^{-i\gamma}$, where both $\rho$
and $\gamma$ are time-dependent\footnote{Here $\gamma$ satisfies the differential equation
\bd
  \frac{d\gamma}{dt}=\frac{\text{const}}{M\rho^2}
\ed
where const is a constant of integration.}
with initial conditions\footnote{Here we note that the initial 
conditions given in (\ref{a IC}) are those that minimize the instantaneous 
energy density for each mode $\vk$ at the initial time $t=t_0$. The states correspond to the 
``instantaneous positive frequency solution," hence the solution (\ref{phi t EOM}) define the 
instantaneous adiabatic ground state \cite{Jacobson:2003vx}.}
\be
  q_k(t=t_0)=\frac{e^{-i\gamma_0}}{\sqrt{M_0\omega_0}}, \hs \text{and} \hs \frac{dq_k}{dt}(t=t_0)=-i\sqrt{M_0\omega_0}e^{-i\gamma_0}=-iM_0\omega_0q_k(t=t_0)
  \label{a IC}
\ee
where $\gamma(t=t_0)\equiv\gamma_0$ is an undetermined initial condition.\footnote{To be 
consistent with FSF we must choose that $\gamma_0=0$ and the constant is equal to one.} 
From (\ref{Ham}), the time-dependent amplitude of the field satisfies the equation of 
motion
\be
  \frac{d^2q_k}{dt^2}+\sigma\frac{dq_k}{dt}+\omega^2q_k=0
  \label{phi t EOM}
\ee
Therefore one can instead solve the linear equation (\ref{phi t EOM})
with initial conditions 
(\ref{a IC}) and use the fact that $\rho=|q_k|$. 

As we have seen in this section, the Bogolyubov transformation between the harmonic 
oscillator and the invariant operator Fock space representations boils down to solving the 
auxiliary equation for $\rho$. The asymptotics of the occupation number given in (\ref{N}) 
will also depend on the asymptotics of $\rho$.\footnote{Note that we could have written the 
occupation number in terms of the amplitudes as
\bd
  N=\frac{M}{2\omega}\left(\left|\frac{dq_k}{dt}\right|^2+\omega^2|q_k|^2\right)-1,
\ed
where upon using the Ermakov equation yields twice the occupation number found in (\ref{N}). It is interesting 
to note that using the occupation number derived from the amplitudes, $|\zeta_k|^2$ now becomes
\bd
  |\zeta_k|^2=\frac{M^2\rho^2\dot\rho^2+M^2\omega^2\rho^4+1}{M^2\rho^2\dot\rho^2+(M\omega\rho^2-1)^2}
\ed
which is of slightly different form than that found using the invariant operator, namely the numerator takes on 
a different form since there is no longer a cross-term.}

\subsection{Bogolyubov Method}

Alternatively one can use the more traditional method, which is to decompose the field in terms of 
different sets of basis functions, $\{f_k(\vec x)\}$ and $\{g_k(\vec x)\}$, and creation and annihilation operators, $\phi_k$ and $\varphi_k$:
\begin{align*}
  \Phi(x)&=\sum_{\vec k}\left(\phi_{\vec k}(t)f_{\vec k}(\vec x)+\phi_{\vec k}^\dagger(t) f^*_{\vec k}(\vec x)\right)\\
     &=\sum_{\vec k}\left(\varphi_{\vec k}(t)g_{\vec k}(\vec x)+\varphi_{\vec k}^\dagger(t) g^*_{\vec k}(\vec x)\right).
\end{align*}
The transformation connecting the two sets of modes $\{f_{\vec k}\}$ and $\{g_{\vec k}\}$ is then 
given by
\bd
  f_{\vec k}(\vec x)=\sum_{\vec k'}\left(\alpha_{\vec k\vec k'}g_{\vec k'}(\vec x)+\beta_{\vec k\vec k'}g^*_{\vec k'}(\vec x)\right),
\ed
which is the Bogolyubov transformation and $\alpha_{\vec k\vec k'}$, $\beta_{\vec k\vec k'}$ are the 
Bogolyubov coefficients. The creation and annihilation operators are connected by
\begin{align*}
  \phi_{\vec k}(t)&=\sum_{\vec k'}\left(\alpha^*_{\vec k\vec k'}\varphi_{\vec k'}(t)-\beta^*_{\vec k\vec k'}\varphi^\dagger_{\vec k'}(t)\right)\nonumber\\
  \varphi_{\vec k'}(t)&=\sum_{\vec k}\left(\alpha_{\vec k\vec k'}\phi_{\vec k}(t)+\beta_{\vec k\vec k'}\phi^\dagger_{\vec k}(t)\right)
\end{align*}

An observer in the $|0_f\rangle$ vacuum, where according to this observer there are no $f$ 
particles, would then observe the average number of $g$ particles
\be
  N_{g_{\vec k}}\equiv\langle0_f|N_{g_{\vec k}}|0_f\rangle=\sum_{\vec k'}|\beta_{\vec k\vec k'}(t)|^2.
  \label{N chi_0}
\ee
If one assumes that asymptotically the spacetime is Minkowski and denote two complex set of 
modes $\phi_{\vec k}^{in}(t)$ and $\phi_{\vec k}^{out}(t)$. Imposing orthogonality, the Bogolyubov 
coefficients become
\be
  \alpha_{\vec k}=-i\left(\phi^*_{\vec k}(t)\dot\varphi_{\vec k}(t)-\varphi_{\vec k}(t)\dot\phi^*_{\vec k}(t)\right)
  \label{alpha_k}
\ee
and
\be
  \beta_{\vec k}(t)=i\left(\phi_{\vec k}(t)\dot \varphi_{\vec k}(t)-\varphi_{\vec k}(t)\dot \phi_{\vec k}(t)\right).
  \label{beta k}
\ee

As we did in the previous section, we can define the number of induced quasi-particles as
\be
  N_{g_{\vec k}}=\frac{1}{|\zeta_{g_{\vec k}}|^2-1},
  \label{op N_0}
\ee
where 
\be
  \zeta_{g_{\vec k}}=\frac{\alpha_k}{\beta_k}=\frac{\varphi_{\vec k}(t)\dot\phi^*_{\vec k}(t)-\phi^*_{\vec k}(t)\dot\varphi_{\vec k}(t)}{\phi_{\vec k}(t)\dot \varphi_{\vec k}(t)-\varphi_{\vec k}(t)\dot \phi_{\vec k}(t)}
  \label{zeta_0}
\ee
where we used (\ref{alpha_k}) and (\ref{beta k}). 

\subsection{2-pt Correlation Function and the Detector Method}

An alternative method for determining the number of induced quasi-particles is to use the 
detector formalism, which calculates the un-equal time 2-point correlation function, also 
known as the Wightman Green function. 

The concept of a detector is to make things more physical, that an observer will have a detector 
that will ``detect" something. The observer knows when the detector has measured something 
when it ``clicks", i.e.~makes a transition from one state to another. Therefore the detector then 
interacts with a field, which causes the detector to make a transition from an initial state, say the 
ground state, to an excited state. 

The simplest detector \cite{Birrell:1982ix}, known as the Unruh-De Witt detector, consists of a 
monopole moment $\alpha(\tau)$ coupled to a scalar field $\Phi(x(\tau))$ at the position of the 
detector at time $\tau$. Here $\tau$ is the propertime of the observer and $x(\tau)$ is the world 
line of the detector as a function of the propertime. The complete Hamiltonian of the system 
\cite{Schlicht:2003iy} is then $H=H_\varsigma+H_\Phi+H_I$, where $H_\varsigma$ is the Hamiltonian 
for the monopole, $H_\Phi$ is the Hamiltonian for the scalar field and $H_I$ is the interaction 
hamiltonian. We will take the interaction Hamiltonian to be
\bd
  H_I=\lambda\varsigma(\tau)\Phi(x(\tau))
\ed
where $\lambda$ is a coupling constant. 

Due to the interaction term, it is easiest to work in the interaction picture, where both states 
and operators are time-dependent. The time evolution of the the states is governed by the 
Schr\"odinger equation
\bd
  H_I|\Phi(\tau)\rangle=\varsigma(\tau)\Phi(x(\tau))|\Psi(\tau)\rangle=i\frac{d}{d\tau}|\Psi(\tau)\rangle
\ed
where $|\Psi(\tau)\rangle$ is taken to be a product state and 
\bd
  \varsigma(\tau)=e^{iH_\varsigma\tau}\varsigma(0)e^{-iH_\varsigma\tau}.
\ed

One can then ask, what is the transition amplitude for finding the detector in an excited state 
$|E,\psi\rangle=|E\rangle\otimes|\psi\rangle$, where $E$ is the energy of the detector at some 
time $\tau$? We can then write, to first order in perturbation theory
\begin{align*}
  A_{i\to f}&=-i\lambda\langle E,\psi|\Psi(\tau)\rangle\nonumber\\
                 &=-i\lambda\langle E|\varsigma(0)|E_0\rangle\int_{\tau_0}^{\tau} d\tau'e^{-i\tau'(E-E_0)}\langle\psi|\Phi(\tau')|\Psi(\tau_0)\rangle
\end{align*}
where $|\Psi(\tau_0)\rangle\equiv|0\rangle$ is the ground state wavefunction. The probability for the transition is then
\begin{align}
  P_{i\to f}&=|A_{i\to f}|^2\nonumber\\
                 &=\lambda^2|\langle E|\varsigma(0)|E_0\rangle|^2\left|\int_{\tau_0}^{\tau} d\tau'e^{-i\tau'(E-E_0)}\langle\psi|\Phi(\tau')|0\rangle\right|^2\nonumber\\
                 &\equiv\lambda^2|\langle E|\varsigma(0)|E_0\rangle|^2{\cal{F}}(E).
  \label{Prob}
\end{align}
From (\ref{Prob}) we can see that the first term only depends on the internal details of the detector, hence we can ignore that part, the relevant part is then ${\cal{F}}(E)$, which is now given by
\begin{align}
  {\cal{F}}(E)&=\int_{\tau_0}^{\tau}d\tau'\int_{\tau_0}^{\tau}d\tau''e^{-iE(\tau'-\tau'')}\langle0|\Phi(\tau')\Phi(\tau'')|0\rangle\nonumber\\
                     &\equiv\int_{\tau_0}^{\tau}d\tau'\int_{\tau_0}^{\tau}d\tau''e^{-iE(\tau'-\tau'')}\langle\Phi(\tau')\Phi(\tau'')\rangle,
  \label{F1}
\end{align}
where the last term $\langle0|\Phi(\tau')\Phi(\tau'')|0\rangle$ is the Wightman function. 
In essence (\ref{F1}) gives the probability of finding the detector in an excited state of energy $E$ above its ground state. However, one can introduce new variables \cite{Schlicht:2003iy} and write (\ref{F1}) as
\be
  {\cal{F}}(E)=\lim_{\epsilon\to0}2\int_{\tau_0}^{\tau}du\int_0^{u-\tau_0}ds\Re\left(e^{-iEs}\langle\Phi(u)\Phi(u-s)\rangle\right).
  \label{F2}
\ee
Finally, one can define the transition rate $\dot{{\cal{F}}}(E)$, which is proportional to the number 
of clicks per second in a detector consisting of identical detector atoms by taking the $u$ 
derivative of (\ref{F2}), which is known as the response function,
\be
  \dot{{\cal{F}}}(E)=\lim_{\epsilon\to0}2\int_0^{\tau-\tau_0}ds\Re\left(e^{-iEs}\langle\Phi(\tau)\Phi(\tau-s)\rangle\right).
  \label{Response}
\ee
Therefore to calculate the response function, one must first calculate the Wightman function. 

To calculate the Wightman function, one expands the scalar field $\Phi(x(\tau))$ as
\be
  \Phi=\int \frac{d^{3}k}{(2\pi)^{3}}\left(\hat a_k\phi_k(\tau)f_k(x)+c.c\right).
  \label{mode}
\ee
Here, the creation and annihilation operators are time-independent and they are defined by
\be
  \hat a_k|0\rangle=0.
  \label{Wightman a}
\ee
Substituting this into the Wightman function leads to
\be
  \langle\Phi(x(\tau))\Phi(x(\tau'))\rangle=\int\frac{d^{3}k}{(2\pi)^{3}}\phi_k(\tau)\phi^*_k(\tau')f_k(x)f_k^*(x').
  \label{Wightman}
\ee
We can see that (\ref{Wightman}) only depends on the mode and basis functions, not on the 
occupation given in (\ref{N}) since (\ref{Wightman}) only depends on the action of the 
annihilation operators on the ground state (\ref{Wightman a}). Also, since $e^{-iEs}$ is entire, 
the Fourier transform only depends on the poles of the Wightman function. 

Notice that for the detector method, (\ref{Wightman}) and (\ref{Response}) are independent of 
the time-dependent Bogoluibov transformation that was central for determining the occupation 
number for the instantaneous diagonalization case. (\ref{Wightman}) only depends on how the 
original annihilation operator acts on the vacuum of the space-time itself, (\ref{Wightman a}).

\section{Friedman-Robertson-Walker space-time}
\label{sec:FRW}

As an application of this, we will consider the standard Friedman-Robertson-Walker (FRW) 
space-time in flat $4$-dimensional space-time. The Lagrangian for the real scalar field propagating in 
the curved space-time background given by the metric tensor $g_{\mu\nu}$ is
\be
  {\cal{L}}=-\frac{1}{2}\left(g^{\mu\nu}\partial_{\mu}\Phi\partial_{\nu}\Phi+\xi R\Phi^2+m^2\Phi^2\right)
  \label{Lag}
\ee
where the metric is $ds^2=-dt^2+a^2(t)d\vx\cdot d\vx$, $a(t)$ is the scale factor,
and $R$ is the Ricci scalar in $t$-coordinates, while the metric is given by 
$ds^2=a^2(\eta)(-d\eta^2+d\vx\cdot d\vx)$ in term so conformal time. 
Decomposing the scalar field into a complete set of 
basis functions denoted by $\{f_k(\vx)\}$, see \cite{Bertoni:1997qb}, as in (\ref{mode}) with 
$\tau$ replaced by $t$ or $\eta$ depending on which coordinates we are woking in, 
it is easy to see that the basis functions take the form $f(\vx)\sim e^{i\vk\cdot\vx}$.

With definition (\ref{mode}), the action for the scalar field can be found from the time integral of 
the Lagrangian (\ref{Lag}). In $t$-coordinates the action for the scalar field is given by
\be
\label{a1}
  S=\int dt\int \frac{d^{3}k}{(2\pi)^{3}}\sqrt{-g}\left[-\dot{\phi}_k^2+\omega_{ph}^2(t)\phi_k^2\right]
\ee
or using $dt=ad\eta$,then in $\eta$-coordinates the action for the scalar field is given by:
\be
\label{a2}
  S=\int d\eta\int \frac{d^{3}k}{(2\pi)^{3}}\frac{\sqrt{-g}}{a}\left[-\left(\frac{d\phi_k}{d\eta}\right)^2+\omega^2(\eta)\phi_k^2\right]
\ee
where, 
\be
  \sqrt{-g}=a^3
  \label{sqrt g}
\ee
and we have defined 
\be
  \omega^2(\eta)\equiv a^2\omega_{ph}^2(t)=k^2+a^2(\eta)(m^2+\xi R)=k^2+a^2(\eta)(m^2+12\xi H^2)\equiv k^2+a^2A^2.
  \label{omega_eta}
\ee
(\ref{omega_eta}) shows that for a massless, minimally coupled scalar field the frequency is 
time-independent. Note that $\omega_{ph}(t)$ is redshifting while $\omega(\eta)$ is blueshifting.

Here we see that written in terms of conformal time\footnote{Note that in terms of propertime 
$t$, $M=\sqrt{-g}$. This will be important when we substitute into (\ref{N}) we will need to use 
the propertime since we are interested in the occupation number as observed by this 
observer.}, $M=\sqrt{-g}/a$ so that using (\ref{sqrt g}) 
we find $\sigma=2a'/a$, which using (\ref{phi t EOM}) leads to the equation of motion
\be
  \frac{d^2\phi_k}{d\eta^2}+2\frac{a'}{a}\frac{d\phi_k}{d\eta}+\omega^2(\eta)\phi_k=0,
  \label{phi_eq}
\ee 
or in terms of $\chi_k=a\phi_k$
\be
  \frac{d^2\chi_k}{d\eta^2}+\tilde\omega^2(\eta)\chi_k=0
  \label{chi_eta}
\ee
where 
\begin{align}
  \tilde\omega^2(\eta)=&k^2+a^2(\eta)A^2-\frac{a''}{a}=\omega^2(\eta)-\frac{a''}{a}.
  \label{omega_pp}
\end{align}

We note that in terms of the $t$-coordinate, we can make a similar replacement so that the frequency of the 
induced quasi-particle is given by
\be
  \tilde\omega^2(t)=\frac{k^2}{a^2}-\frac{3}{2}\left(\frac{\ddot a}{a}+\frac{1}{2}\left(\frac{\dot a}{a}\right)^2\right)+A^2=\omega^2(t)-\frac{3}{2}\left(\frac{\ddot a}{a}+\frac{1}{2}\left(\frac{\dot a}{a}\right)^2\right).
  \label{omega_p}
\ee
Hence, as we can see from (\ref{omega_pp}) and (\ref{omega_p}), $\tilde\omega^2(\eta)$ is 
related to $\tilde\omega^2(t)$ in $3D$, not in $4D$. That is, working 
in conformal time effectively reduces the dimensionality of the problem from $4D$ down to 
$3D$.

In the next section we note that we can either work with the $\phi_k$ or $\chi_k$ fields, 
since as we see from (\ref{N def}) that the occupation only depends on the expectation value 
of the $\hat A_k$ particles in the vacuum $|0\rangle_c$. In the literature, however, it is customary 
to use the $\chi_k$ field to determine the particle production, so we shall concentrate only on 
this field. 

\section{Particle Production in de Sitter Space-Time}
\label{sec:dS prod}

In this section we will consider the particle production which occurs during de Sitter 
expansion using the methods of simultaneous diagonalization, usual Bogolyubov transformations 
between two different basis functions (where the second vacuum is chosen to be the so-called 
adiabatic vacuum), and the Unruh-De Witt detectors.

\subsection{Instantaneous diagonalization}
\label{sec:Instant}

To determine the particle production we need to solve (\ref{chi_eta}) for the $\chi_k$ 
field and use this to find $\rho$. In this section we will specialize to the case of de Sitter expansion.

The solution to the Einstein's equations with a positive cosmological constant give the expansion 
factor $a(t)\sim \exp(Ht)$, where {H} is the Hubble parameter given at late times. 
In terms of the conformal time the scale factor takes the form $a\sim-(\eta H)^{-1}\equiv z^{-1}$.

From the scale factor we can write (\ref{omega_pp}), using that in $4D$ the Ricci scalar is 
$12H^2$, as
\begin{align}
  \tilde\omega^2(\eta)&=k^2+\frac{m^2+2(6\xi-1)H^2}{z^2}\nonumber\\
     &\equiv k^2+\left(\frac{B}{z}\right)^2.
     \label{omega_tilde}
\end{align}
Unlike (\ref{omega_pp}), (\ref{omega_tilde}) shows that the frequency in the case of a massless, 
minimally coupled scalar field is time-dependent, since here $A^2=0$, $B^2=-H^2$. 
We can now use (\ref{chi_eta}) to find the general solution for $\chi_k$, which is given by
\be
  \chi_k=\sqrt{z}\left[\tilde\alpha_kH_y^{(1)}\left(\frac{kz}{H}\right)+\tilde\beta_kH_y^{(2)}\left(\frac{kz}{H}\right)\right],
  \label{chi_sol}
\ee
where
\be
  y\equiv\sqrt{\frac{1}{4}-\left(\frac{B}{H}\right)^2}=\sqrt{\left(\frac{3}{2}\right)^2-\left(\frac{A}{H}\right)^2},
  \label{y}
\ee
which gives the expression for $\phi_k$ given by
\be
  \phi_k=z^{3/2}\left[\tilde\alpha_kH_y^{(1)}\left(\frac{kz}{H}\right)+\tilde\beta_kH_y^{(2)}\left(\frac{kz}{H}\right)\right].
  \label{phi_sol}
\ee
To determine the coefficients $\alpha_k$ and $\beta_k$ we can use (\ref{chi_sol}).

Expanding (\ref{chi_sol}) for large $Z$ and using (\ref{a IC}) we find that the coefficients satisfy
\be
  \tilde\alpha_k=e^{-i\theta}\left(B-\tilde\beta_k e^{-i\theta}\right), \hs \tilde\beta_k=-\frac{i}{k}\frac{2BH}{z_0}e^{i\theta}
  \label{alpha_beta}
\ee
where we have defined
\be
  B\equiv\sqrt{\frac{\pi kH}{2z_0}}\phi(z_0)=\sqrt{\frac{H\pi}{2z_0}},
  \label{B}
\ee
and 
\bd
  \theta\equiv \frac{kz_0}{H}-y\frac{\pi}{2}-\frac{\pi}{4}.
\ed
Using (\ref{alpha_beta}) and (\ref{B}), we can rewrite $\alpha_k$ as
\bd
  \tilde\alpha_k=e^{-i\theta}B\left(1+\frac{2iH}{kz_0}\right).
\ed
It is customary to compare these terms and keep only the $\beta_k$ term\footnote{This is 
due to the fact that the $\beta_k$ cancels the $\eta_0$ dependence of the coefficient in 
(\ref{chi_sol}).}, which leads to the 
usual Bunch-Davies vacuum, hence we are left with\footnote{Alternatively, one can restrict the initial 
vacuum to consist of only the positive frequency modes, meaning that one should only keep the 
$\tilde\beta_k$ term. Expanding the Hankel function for large argument one then easily arrives at 
(\ref{chi_Bunch}).}
\be
  \chi_k(\eta)=e^{i\theta}\sqrt{\frac{\pi z}{2H}}H^{(2)}_y\left(\frac{kz}{H}\right),
  \label{chi_Bunch}
\ee
so that (\ref{phi_sol}) becomes
\be
  \phi_k=e^{i\theta}\sqrt{\frac{\pi}{2H}}z^{3/2}H^{(2)}_y\left(\frac{kz}{H}\right).
  \label{phi_Bunch}
\ee

From (\ref{chi_Bunch}) we see that the auxiliary field $\rho$ is given by
\be
  \rho=\sqrt{\frac{\pi ze^{\pi\Im y}}{2H}}\sqrt{H_{y^*}^{(1)}\left(\frac{kz}{H}\right)H_y^{(2)}\left(\frac{kz}{H}\right)},
  \label{rho_ex}
\ee
and from this we have
\begin{align}
  \rho_z&=\frac{d}{dz}\sqrt{\chi_k\chi_k^*}=\frac{1}{2\rho}\left(\frac{d\chi_k}{dz}\chi_k^*+\chi_k\frac{d\chi^*_k}{dz}\right)\nonumber\\
    &=\frac{\rho}{2}\left[\frac{1}{z}+\frac{k\pi z e^{\pi\Im(y)}(H_y^{(2)}(x)(H_{y^*}^{(1)}(x))'+H_{y^*}^{(1)}(x)(H_{y}^{(2)}(x))')}{2H^2\rho^2}\right]
    \label{drho_ex}
\end{align}
where the prime denotes differentiation with respect to the $x$, which is defined as $x=kz/H$. 
From (\ref{omega_tilde}), (\ref{rho_ex}), and (\ref{drho_ex}) we can then solve for the occupation 
number exactly as a function of $\eta$.\footnote{Since we are using the $\chi_k$ field, we see that 
in this case $M=$const$=1$.} 

We will be mostly interested in the asymptotic behavior of the occupation number given in (\ref{N}). 
We will be interested in different limits: The first is the more natural limit, when we take the limit 
that $z\to0$, and the second limit we are interested in the large $k$ portion of the spectrum; 
that is in the limit where $k$ is large and $z$ is some fixed finite time. 
For the first limit we take a fixed time $z$, and consider the case where in (\ref{omega_tilde}) 
$k<<A/z$, or $z<<A/k$, thus we have that $\omega(\eta)\approx A/z$. Furthermore, we are interested 
in the case of a very massive field, that is $m>>H$, which is forced upon us by the requirement 
of the adiabatic vacuum in Section \ref{sec:Adiabatic}. In this limit, we can see from (\ref{omega_tilde}) and 
(\ref{omega_eta}) that $A\approx B$, hence in this limit 
$\omega\approx\tilde\omega$. In Appendix \ref{second case} we will be interested in the opposite limit, that 
is when $m<<H$, however as we will see in from Section \ref{sec:Adiabatic} this is outside of the regime 
of validity for the adiabatic approximation.


For $m>>H$, from (\ref{y}) $y=iA/H$, and we can expand the Hankel functions in the $z\to0$ limit: This limit 
corresponds to $A$ being large, or that we are deep in the UV (small distance) limit,
\be
  \chi_k\approx e^{i\theta'}\sqrt{\frac{zHe^{\pi A/H}}{2\pi A^2}}\left[e^{-\pi A/H}\Gamma^*\left(1+i\frac{A}{H}\right)\left(\frac{kz}{2H}\right)^{iA/H}-\Gamma\left(1+i\frac{A}{H}\right)\left(\frac{kz}{2H}\right)^{-iA/H}\right]
\ee
where 
\be
  \theta'\equiv k|\eta_0|-\frac{\pi}{4},
\ee
which gives 
\begin{align}
  \rho&=\sqrt{\frac{z}{A\tanh(\pi A/H)}}=\sqrt{\frac{1}{\omega\tanh(\pi\omega z/H)}}.
  \label{late rho}
\end{align}
Additionally, one can check that from that (\ref{drho_ex}) $\rho_z\approx\rho/(2z)$, so we can 
find that the occupation of induced quasi particles for fixed $\eta$ is given by
\be
  N\approx\frac{1}{4\tanh(\pi\omega z/H)}\left[\left(1-\tanh(\pi\omega z/H)\right)^2+\left(\frac{1}{2\omega z}\right)^2\right]
\ee
or for large $\tilde\omega$ (or for large $A$) we have
\be
  N\to\frac{1}{16\omega^2z^2}=\frac{1}{16A^2}=\frac{H^2}{16\omega_{ph}^2},
  \label{Napprox}
\ee
where we defined the physical frequency as $\omega_{ph}=\omega/a$, which is defined in terms of 
the original frequency (\ref{omega_eta}). (\ref{Napprox}) corresponds to a nonthermal, polynomial 
distribution for the occupation number.\footnote{Notice that calculating the occupation number in terms 
of $\omega(\eta)$, that is in terms of conformal time, is equivalent to calculating the occupation number 
in terms of $\omega(t)$, that is in terms of the flat coordinate time $t$. This is because in $t$, $M=a^3$ 
and $\rho_\phi=\rho_\chi/a^{3/2}$, hence $M\rho_\phi^2=\rho_\chi$, and additionally 
$\omega(t)=\omega(\eta)/a=A$, which is just a constant at late times.}

In the late time approximation (\ref{zeta}), with the help of (\ref{late rho}) and 
$\rho_\eta\approx\rho/(2\eta)$, becomes
\bd
  \zeta\approx\frac{1+\rho^2(\tilde\omega+i/(2\eta))}{1-\rho^2(\tilde\omega+i/(2\eta))}=\frac{1+\tilde\omega\rho^2(1+i/(2A))}{1-\tilde\omega\rho^2(1+i/(2A))}=\frac{\tanh(\pi A)+(1+i/(2A))}{\tanh(\pi A)-(1+i/(2A))}.
\ed
Substituting into (\ref{op N}) yields
\bd
  N=\frac{1+4A^2(\tanh(\pi A/H)-1)^2}{4A^2((\tanh(\pi A/H)+1)^2-(\tanh(\pi A/H)-1)^2)}
\ed
or in the large $A$ limit
\be
  N\to\frac{1}{16A^2}=\frac{H^2}{16\omega_{ph}^2}.
  \label{op N large}
\ee
Comparing (\ref{op N large}) to (\ref{Napprox}), we see that this has the same 
structure and is thus consistent with our previous results.\footnote{Here we note that using the occupation 
number derived from the amplitudes, we find that the occupation number is given by
\bd
  N\approx\frac{1}{8A^2}=\frac{H^2}{8\omega^2}
\ed
which is consistent with our conclusion that the occupation derived from the amplitudes is twice that of the 
occupation number derived from the invariant operator.}

Now consider for fixed finite time $z$, the behavior of the tail, that is the asymptotic behavior 
of the occupation number. In this limit, we take that $k>>A/z$, so that $\omega\approx k$ 
and $\rho\approx1/\sqrt{k}$, so that the occupation number can be written as
\be
  N\to\frac{1}{4k^2z^2}=\frac{1}{4\omega^2z^2}=\frac{H^2}{4\omega_{ph}^2}
  \label{Napproxk}
\ee
which again is a non-thermal, polynomial behavior. At this point one can of course worry about our 
assumption that $k>>A/H\eta$, since these two factors are of the same order as $\eta$ goes to zero. 



\subsection{Bogolyubov Method: Adiabatic Vacuum}
\label{sec:Adiabatic}

In this section we shall consider the usual Bogolyubov transformation between two distinct 
vacua, where the second vacuum is chosen to be the so-called adiabatic vacuum, which is 
given by
\be
  \chi_{m,k}=\frac{1}{\sqrt{2W_m}}e^{-i\int_{z_0}^z dz'W_m/H}
  \label{chi_m}
\ee
where $\tau$ stands for either $t$ or $\eta$, and $W_m$ satisfies the nonlinear equation
\be
  W^2_m=\tilde\omega^2-\frac{1}{2}\left(\frac{(W_m)_{zz}}{W_m}-\frac{3}{2}\frac{(W_m)_{z}^2}{W_m^2}\right).
  \label{W_m}
\ee
If the spacetime is slowly varying, hence if the frequency $W_m$ is slowly varying, then the derivative 
terms in (\ref{W_m}) will be small compared to $\tilde\omega^2$, thus we can take the zeroth 
order limit of (\ref{chi_m}), which yields
\be
  \chi_{0,k}=\frac{1}{\sqrt{2\omega}}e^{-i\int_{z_0}^z dz'\tilde\omega(z')/H}.
  \label{chi_0}
\ee
A quantitative condition for $\omega$ to be a slowly-changing function of $z$ is that the relative 
change of $\omega$ during one oscillation period is negligibly small, hence we must require
\be
  \left|H\frac{\tilde\omega_z}{\tilde\omega^2}\right|<<1.
  \label{adiab cond}
\ee

For our purposes, from (\ref{adiab cond}), this means that we must have that 
\bd
  \left|H\frac{\tilde\omega_z}{\tilde\omega^2}\right|=\frac{HB^2}{(k^2z^2+B^2)^{3/2}}=\frac{H}{B((kz/B)^2+1)^{3/2}}
\ed
so that for $z\to\infty$ this is satisfied and for $z\to0$ we have that
\bd
  1<<\frac{B}{H}=\frac{1}{H}\sqrt{A^2-2H^2},
\ed
which means that $A>>H$, which corresponds to the first limit in the previous section. 
Thus as $a\to\infty$, unless $H<<1$, only the largest modes are valid in the adiabatic 
approximation. In terms of de-Sitter space, these modes correspond to those well inside of the 
horizon approaching the limit where the modes can be treated as if they are in pure Minkowski 
space. 

Using (\ref{omega_tilde}) we can rewrite (\ref{chi_0}) as 
\be
  \chi_{0,k}=\frac{1}{\sqrt{2\tilde\omega}}e^{i(z_0\tilde\omega_0-z\tilde\omega)}\left[\frac{B+z_0\tilde\omega_0}{B+z\tilde\omega}\right]^{-iB/H}\left|\frac{z}{z_0}\right|^{-iB/H}.
  \label{Chi_0}
\ee
It is easy to check that (\ref{Chi_0}) has the same initial condition as (\ref{a IC}), with $\gamma_0=0$:
\bd
  \chi(z=z_0)=\chi_0(z=z_0). 
\ed
For the time-derivative of (\ref{Chi_0}) we have
\begin{align}
  \frac{d\chi_{0,k}}{dz}=\chi_{0,k}\left[\frac{B^2}{2z^3\omega^2}-i\frac{\tilde\omega}{H}\right].
      \label{dchi_0}
\end{align}
At $z=z_0$, (\ref{dchi_0}) becomes
\begin{align}
  \frac{d\chi_{0,k}}{dz}(z=z_0)\approx -i\frac{\tilde\omega_0}{H}\chi_{0,k}(z=z_0).
     \label{dPhi_0}
\end{align}
Here, we see that (\ref{dPhi_0}) is the same as the initial condition in (\ref{a IC}) 
again with $\gamma_0=0$. 

To calculate the occupation number in the adiabatic vacuum, one can no longer use (\ref{N}) since 
the adiabatic vacuum and the invariant operator method are no longer related to each other by 
the time-dependent Bogolyubov transformation in (\ref{bog coeff}). Therefore, one must use 
(\ref{N chi_0}) and (\ref{beta k}) to determine the occupation number. If we 
use the zero order adiabatic vacuum, (\ref{N chi_0}) becomes 
\be
  N_{\chi_0}=|\beta_{\chi_0}|^2=\left|\chi_k\frac{d\chi_{0,k}}{dz}-\frac{d\chi_k}{dz}\chi_{0,k}\right|^2=\left|\chi_k\chi_{0,k}\right|^2\left|\frac{B^2}{2z^3\tilde\omega^2}-i\frac{\tilde\omega}{H}-\frac{1}{2z}-\frac{k}{H}\frac{(H^{(2)}_y(x))'}{H_y^{(2)}(x)}\right|^2. 
  \label{N_0}
\ee
From (\ref{Chi_0}) we see that $|\chi_{0,k}^i|^2=(2\omega)^{-1}$, so we can write 
\be
  N_{\chi_0}=\frac{\rho^2}{2\omega}\left|\frac{B^2}{2z^3\tilde\omega^2}-i\frac{\tilde\omega}{H}-\frac{1}{2z}-\frac{k}{H}\frac{(H^{(2)}_y(x))'}{H_y^{(2)}(x)}\right|^2
  \label{N_02}
\ee
Using again $\tilde\omega\approx B/z\approx A/z$ and (\ref{Chi_0}), (\ref{N_0}) (or (\ref{N_02})) becomes
\begin{align}
  N_{\chi_0}&=\frac{\rho^2}{2\tilde\omega}\left|i\frac{A}{z}+\frac{k}{H}\frac{(H^{(2)}_y(x))'}{H_y^{(2)}(x)}\right|^2\nonumber\\
    &=\frac{\pi e^{\pi A}}{4z\tilde\omega}\left|iAH_y^{(2)}(x)+x(H_y^{(2)}(x))'\right|^2\nonumber\\
    &=\frac{\pi e^{\pi A}}{4z\tilde\omega}\left|-\frac{2i}{\pi}e^{-\pi A}\Gamma(1-iA)\left(\frac{x}{2}\right)^{iA}\right|^2\nonumber\\
    &=\frac{e^{-\pi A}}{\pi z\tilde\omega}\left|\Gamma(1-iA)\right|^2=\frac{1}{e^{2\pi\omega_{ph}/H}-1},
    \label{N_0f}
\end{align}
which is the thermal distribution with temperature
\be
  T=\frac{1}{2\pi\eta}=\frac{H}{2\pi}=T_H
  \label{Temp Bog}
\ee
which is divergent as $\eta\to0$ for the comoving observer but gives the expected temperature 
for the physical observer. Interestingly, when comparing to the standard result of the response 
function of an Unruh-de Witt detector, (\ref{N_0f}) corresponds to a thermal distribution in $3D$ 
not the full $4D$ in which we are working\footnote{This is because in $4D$ the thermal distribution 
goes as \cite{Casadio:2010vq,Garbrecht:2004du}
\bd
  N\sim\frac{\omega}{e^{\beta\omega}-1}
\ed
and in $3D$ the thermal distribution goes as
\bd
  N\sim\frac{1}{e^{\beta\omega}-1}.
\ed
Hence the thermal distribution in (\ref{N_0f}) corresponds to the $2D$ case, at least in the context 
of the Unruh-de Witt detectors.}, which makes sense according to our above discussion on how 
working in conformal coordinates effectively decreases the dimensionality by one. 

Alternatively, using (\ref{zeta_0}) we have
\begin{align}
  \zeta_0=i\frac{H_{y^*}^{(1)}(x)}{H_y^{(2)}(x)}e^{\pi A},
\end{align}
where upon taking the absolute value the coefficient becomes one, since 
$(H_y^{(2)}(x))^*=H^{(1)}_{y^*}(x)$, and we are left with 
\bd
  |\zeta_0|^2=e^{2\pi A}=e^{2\pi\omega_{ph}/H}.
\ed
From (\ref{op N_0}) we then have
\be
  N_{\chi_0}=\frac{1}{e^{2\pi\omega_{ph}/H}-1}
\ee
which is the same as in (\ref{N_0f}) as expected. 

As a side note, notice that if we use (\ref{Chi_0}) instead of (\ref{chi_Bunch}) in (\ref{N}) we 
arrive at,\footnote{Here $N_0$ denotes the occupation number (\ref{N}) in respect to $\chi_0$ 
(\ref{Chi_0}).}
\be
  N_0=\frac{1}{4}\left(\frac{A^2}{2\tilde\omega^3\eta^3}\right)^2=\frac{1}{16A^2},
  \label{N0approx}
\ee
where we again used that $\tilde\omega\approx A/\eta$. We can easily see that (\ref{N0approx}) is 
the same as (\ref{Napprox}), which again is a nonthermal, polynomial distribution.

Hence to achieve the thermal distribution, one must use the adiabatic vacuum, that is restrict the 
modes to those satisfying the condition $A>>1$ (that is for $m>H$, which corresponds to the 
non-relativistic modes), and consider the Bogolyubov transformation 
between the instantaneous vacuum, in this the Bunch-Davies vacuum (\ref{chi_Bunch}), and 
the adiabatic vacuum (\ref{Chi_0}). The occupation number given by (\ref{N}), the instantaneous 
number of particles between the Harmonic oscillator vacuum and the invariant operator 
vacuum, as seen from (\ref{Napprox}) and (\ref{N0approx}) generically gives that the spectrum 
of particles is $A^{-2}=(\tilde\omega\eta)^{-2}=\omega_{ph}^{-2}$. 

\subsection{Unruh-de Witt Detector}
\label{sec:Unruh}

Finally we shall consider the response of an Unruh-de Witt detector. 
To determine the Wightman function in (\ref{Wightman}) we use the Bunch-Davies vacuum 
(\ref{chi_Bunch}). Upon substitution into (\ref{Wightman}) yields
\be
  W=\langle\Phi(x(\eta))\Phi(x(\eta'))\rangle
    =\frac{\pi e^{\pi\Im y}}{2Ha^{3/2}(\eta)a^{3/2}(\eta')}\int \frac{d^3k}{(2\pi)^3}H_y^{(2)}(k\eta)H_{y^*}^{(1)}(k\eta')
  \label{Wightman_int}
\ee
where $\Delta\vec x=x(\eta)-x(\eta')$. The conformal coordinate interval, $\bar x$ and the geodesic 
distance, $\ell=\ell(x;x')$, are related by \cite{Garbrecht:2004du,Prokopec:2003tm}
\be
  p=p(x;x')\equiv a(\eta)a(\eta')H^2\Delta\bar x^2(x;x')=4\sin^2\left(\frac{H\ell(x;x')}{2}\right).
  \label{geodesic}
\ee
Here we will consider an observer moving along the geodesic 
$p=p(x(t+\Delta t/2);x(t-\Delta t/2))=-4\sinh^2(H\Delta t/2)$, that is we take $\ell\to i\Delta t$. 

We know that $\Phi(x)$ satisfies (\ref{phi_eq}), however we can see that the two-point Wightman 
function satisfies the same differential equation. For a de Sitter invariant vacuum state, the state can 
only depend on the geodesic distance between $x$ and $x'$, which from (\ref{geodesic}) may be 
written as a function $z(x;x')\equiv1-p/4$ \cite{Garbrecht:2004du,Prokopec:2003tm}. Using this we 
can then rewrite (\ref{phi_eq}) as
\be
  z(1-z)\frac{d^2}{dz^2}W+4\left(\frac{1}{2}-z\right)\frac{d}{dz}W-\left(m^2+\xi R\right)W=0.
  \label{W_eq}
\ee
Note that (\ref{W_eq}) has a $z\leftrightarrow1-z$ symmetry, hence we have the solution for 
$W(p)\equiv W(z)$ \cite{Garbrecht:2004du,Prokopec:2003tm}
\be
  iW(p)=\frac{H^2}{4\pi^2}{}_2F_1\left(\frac{3}{2}-y,\frac{3}{2}+y,2;1-\frac{p}{4}\right)\Gamma\left(\frac{3}{2}-y\right)\Gamma\left(\frac{3}{2}+y\right).
  \label{W_p}
\ee

If we consider the case where $|\xi|<<1$ and $m<<H$, expanding in powers of 
\bd
  \frac{3}{2}-y=\frac{m^2+\xi R}{3H^2},
\ed
one can then find that the response rate (\ref{Response}) is given by, using (\ref{W_p})
\bd
  \dot{\cal F}=\frac{\Delta E}{2\pi}\left(1+\frac{H^2}{\Delta E^2}\right)\frac{1}{e^{2\pi\Delta E/H}-1}+\frac{H^2}{2\pi}\delta(\Delta E)\left(\frac{1}{3-2y}-1+\ln(2)+{\cal O}\left(\frac{3}{2}-y\right)\right).
  \label{small m}
\ed

Of course here we took the opposite limit as we have taken in the previous two cases, that is here 
we have taken that $A<<1$, while in the previous cases we have taken that $A>>1$. To consider this 
case we consider (\ref{Wightman_int}), which in coordinate time $t$ gives the response function 
\cite{Higuchi:1986ww}
\be
  \dot{\cal F}=\frac{H}{4\pi^3}e^{-\pi\Delta E/H}\left|\Gamma\left(\frac{3/2+i\Delta E/H+y}{2}\right)\Gamma\left(\frac{3/2+i\Delta E/H-y}{2}\right)\right|^2.
  \label{R1}
\ee
In the desired limit, $y\approx iA$ and $\omega\approx A\eta=H\eta\omega_{ph}=HA$, so we 
have (with $\Delta E/H=\omega_{ph}/H=A$) from (\ref{R1}), 
\begin{align}
  \dot{\cal F}=\frac{H}{4\pi^3}e^{-\pi A}\left|\Gamma\left(\frac{3}{4}+iA\right)\Gamma\left(\frac{3}{4}\right)\right|^2.
  \label{R2}
\end{align}
(\ref{R2}) has the asymptotic structure given by
\be
  \dot{\cal F}\approx\frac{H}{2\pi^2}\left|\Gamma\left(\frac{3}{4}\right)\right|^2\sqrt Ae^{-2\pi A}=\frac{H}{2\pi^2}\left|\Gamma\left(\frac{3}{4}\right)\right|^2\sqrt{\frac{\omega_{ph}}{H}}e^{-2\pi\omega_{ph}/H}
  \label{R3}
\ee
which approaches a thermal distribution only for large $A=\omega_{ph}/H$. We note that 
(\ref{R3}) has a strange, fractional dimensional dependence.

\section{Discussion}
\label{sec:Discussion}

In this paper we have investigated particle production, induced quasi-particles production, 
during a process 
where the ``mass" and frequency of the particles are time-dependent. First, we considered the 
occupation number of the induced quasi-particles for a general time-dependent system using 
the instantaneous diagonalization method, usual Bogolyubov transformation between two 
different basis states for the same scalar field, and Unruh-de Witt detector method. We 
showed that the occupation number as derived from instantaneous diagonalization of the 
time-dependent Hamiltonian: that is the Functional Schr\"odinger formalism 
(FSF) (under first quantization), quantum kinetic equation (under second quantization), and 
Bogolyubov transformation (under second quantization) between two different Fock 
space representations of the general time-dependent system, all give the same result. The 
general result is that all three of these methods amount to the counting of quasi-particle production 
between the harmonic oscillator basis states and the invariant operator basis states, which 
are different Fock space representations of the time-dependent problem. Second, 
as an example of the different methods, to compare and contrast the results, we considered the 
Friedman-Robertson-Walker (FRW) space-time in $4D$, using the conformal time coordinates. 
Using the FRW space-time, we derived the usual equations of motion for the mode coefficients 
of the scalar field $\Phi$, $\phi_k(\eta)$, and then defined a new field $\chi_k$ to simplify the
equations of motion. Third, we specialized to the example of de Sitter space-time to derive the 
occupation number in the three different methods. 

The results are as follows:
\begin{itemize}
  
  \item In Sections \ref{sec:Instant} and \ref{sec:Adiabatic} we showed that the instantaneous 
  diagonalization method generically yields a nonthermal, polynomial distribution for the 
  occupation number. This was shown using the usual standard Buch-Davies vacuum, 
  (\ref{Napprox}), and the so-called adiabatic vacuum (\ref{Napproxk}). 
  
  \item In Section \ref{sec:Adiabatic} we showed that the standard Bogolyubov transformation 
  method, here we consider the transformation between the instantaneous vacuum, the 
  Bunch-Davies vacuum, and the so-called adiabatic vacuum, in the late time, large $A$, limit 
  the occupation number does follow a thermal distribution (\ref{N_0f}). Interestingly, when 
  comparing with the typical Unruh-de Witt detector result, we see that the obtained thermal 
  distribution in (\ref{N_0f}) is that of the $2D$ instead of the full $4D$ case that we are 
  investigating here. In most cases this side fact is not a problem, since one is only concerned 
  with the appearance of the thermal spectrum, i.e.~the Planck distribution. 
  
  \item From (\ref{dchi_0}) and (\ref{chi_Bunch}) we see that both the Bunch-Davies and 
  adiabatic vacua have the same asymptotic behavior in the late time limit:
  \be
    \frac{d\chi_k}{d\eta}\sim\chi_k(\eta\to0)\left[\frac{1}{\eta}+i\tilde\omega{\cal G}(A)\right]=\chi_k(\eta\to0)\tilde\omega\left[\frac{1}{A}+i{\cal G}(A)\right],
    \label{deriv}
  \ee
  where ${\cal G}(A)$ is a function of $A$. This means that neither of the vacua give rise to the 
  notion of a well-defined particle in the late time limit, which is due to the fact that in the late 
  time limit $\tilde\omega$ diverges. As we can see from (\ref{N_02}), the extra terms that 
  cause the difficulty of defining a well-defined notion of a particle are canceled due to the 
  subtraction of the two derivatives, leading to the thermal spectrum in (\ref{N_0f}). 
  
  \item Furthermore, notice that if $\chi_k$ and $\chi_{0,k}$ satisfy (\ref{a IC}) and (\ref{dPhi_0}) 
  for all time, that is the instantaneous and adiabatic vacua have well-defined notions of 
  particles, then (\ref{N}) naturally predicts that there is particle creation\footnote{Here one 
  can easily check that, from (\ref{drho_ex}), $\rho_{\eta}=0$. However, from (\ref{N}) we see that 
  \bd
    N=\frac{\omega\rho^2}{4}\left(1-\frac{1}{\omega\rho^2}\right)^2
  \ed
  which, unless $\rho^2=1/\omega$, will in general be non-zero. For example, in the definition of 
  the adiabatic vacuum (\ref{chi_m}), we see that $\rho^2\not=1/\omega$ due to the factor of 
  $\sqrt2$ in the denominator.} while (\ref{N_0}) 
  naturally predicts that there is no particle creation since the difference of the two derivatives 
  cancel leaving zero particle creation. This can be avoided if $\chi_k$ and $\chi_{0,k}$ 
  satisfy a condition similar to (\ref{deriv}) where the function ${\cal G}(A)$ for the two vacua 
  are not the same. 
  
  \item In Section \ref{sec:Unruh} we showed that in the large $A$ limit, the spectrum of the 
  Response function, (\ref{R1}), approaches a thermal distribution. However, the distribution in 
  (\ref{R3}) seems to follow a distribution for a fractional dimension instead of the full $4D$ case.  
  
\end{itemize}

The results imply that the instantaneous diagonalization of the Hamiltonian (that is FSF, 
quantum kinetic equation, and the Bogolyubov transformation between the time dependent 
vacua of the harmonic oscillator and the invariant operator) generically give a polynomial 
dependence for the occupation number of the induced quasi particles in the flat FRW chart 
in de Sitter space-time. This behavior comes from the fact that in the late time approximation, 
the derivative of the mode functions in (\ref{deriv}) do not lead to well-defined concepts of 
particles due to the divergence of the first term. In cases where the divergent term is not present, 
it can be shown that the usual Bogolyubov transformation method and the instantaneous 
diagonalization method give the same result, see for example \cite{Haro,Haro:2009zza}. 
The usual Bogolyubov transformation and the 
Unruh-de Witt detector methods both lead to a thermal spectrum in the limit that $A>>1$ 
($\omega_{ph}/H>>1$). One should expect, however, that only those modes that have information about 
the horizon should fit a thermal distribution. The physical frequency is, for early times
\bd
  \omega\approx\frac{k}{a}
\ed
since $a$ goes as $e^{-t}$ or goes to infinity, and for late times we have
\bd
  \omega\approx\sqrt{m^2+\xi R}>>H
\ed
by assumption to satisfy the adiabatic approximation, which is well inside the horizon. Thus the modes 
start well inside the horizon and end well inside the horizon, thus we should expect that the distribution 
is nonthermal. Interestingly, neither of these methods leads to the correct dimensionality 
of the problem; that is neither of the methods leads to a thermal distribution in $4D$.

\appendix
\section{$m<<H$ case}
\label{second case}

From (\ref{omega_tilde}) we see that $(A/H)^2=12\xi$ and from (\ref{y}) we see that $y$ is then a 
real constant: $0<y\leq3/2$. In this case we have
\be
  \chi_k\approx e^{i\theta}\sqrt{\frac{z}{2\pi Hy^2}}\left[e^{-i\pi y}\Gamma\left(1-y\right)\left(\frac{x}{2x}\right)^{y}-\Gamma\left(1+y\right)\left(\frac{x}{2}\right)^{-y}\right]
\ee
so that $\rho$ can be written as
\begin{align}
  \rho&=\sqrt{\frac{z}{2\pi Hy^2}}\sqrt{\Gamma^2(1-y)\left(\frac{x}{2}\right)^{2y}+\Gamma^2(1+y)\left(\frac{2}{x}\right)^{2y}-\frac{2\pi y}{\tan(\pi y)}}\nonumber\\
     &\approx\sqrt{\frac{z}{2\pi Hy^2}}\sqrt{\Gamma^2(1+y)\left(\frac{2}{x}\right)^{2y}-\frac{2\pi y}{\tan(\pi y)}}
\end{align}
and from (\ref{drho_ex}) we have
\begin{align}
  \rho_z&=\frac{\rho}{2}\left[\frac{1}{z}+\frac{1}{\pi Hy\rho^2}\left(\Gamma^2(1-y)\left(\frac{x}{2}\right)^{2y}-\Gamma^2(1+y)\left(\frac{2}{x}\right)^{2y}\right)\right]\nonumber\\
      &\approx\frac{\rho}{2}\left[\frac{1}{z}-\frac{\Gamma^2(1+y)}{\pi Hy\rho^2}\left(\frac{2}{x}\right)^{2y}\right]\nonumber\\
      &=\frac{\rho}{2z}\left[1-\frac{2y}{\rho^2}\left(\rho^2+\frac{2\pi y}{\tan(\pi y)}\right)\right]
\end{align}
Thus the structure of the occupation number is complicated since it depends on the value of $y$ and will 
in general depend on the value of $k$.


As in the previous section, in this case we take $y$ to be a real constant. In this case 
(\ref{N_0}) becomes
\begin{align}
  N_{\chi_0}&=\frac{\rho^2}{2\tilde\omega}\left|i\frac{B}{z}+\frac{k}{H}\frac{(H^{(2)}_y(x))'}{H_y^{(2)}(x)}\right|^2\nonumber\\
    &=\frac{\pi}{4z\tilde\omega}\left|iBH_y^{(2)}(x)+x(H_y^{(2)}(x))'\right|^2\nonumber\\
    &=\frac{1}{z\pi\omega}\left|e^{i\pi y}\Gamma(1-y)\left(\frac{x}{2}\right)^{y}\left(\frac{iB}{y}+1\right)+\Gamma(1+y)\left(\frac{x}{2}\right)^{-y}\left(1-\frac{iB}{y}\right)\right|^2\nonumber\\
    &\approx\frac{1}{\pi z\tilde\omega}\left[4^y(B^2+y^2)\frac{\Gamma^2(y)}{x^{2y}}+\frac{2\pi}{y}((y^2-B^2)\cot(\pi y)-2By)\right]\nonumber\\
    &=\frac{1}{\pi z\omega}\left[4^y\left(\frac{1}{2}\right)^2\frac{\Gamma^2(y)}{x^{2y}}+\frac{2\pi}{y}\left(\left(\frac{1}{4}-2B^2\right)\cot(\pi y)-2By\right)\right]
    \label{N_0fw}
\end{align}
where we used (\ref{y}) in the last line. However, we are well outside the validity of the adiabatic modes so 
we cannot assume that (\ref{N_0fw}) will be valid in this regime.

\subsection{Massless, Conformally Coupled Scalar Field}

In the massless, conformally coupled scalar field we can see from (\ref{omega_tilde}) that 
\bd
  \tilde\omega^2=k^2
\ed
which is constant, hence we shouldn't expect to measure any particles.\footnote{From (\ref{omega_p}), 
this will not be the case in $t$-coordinates, since in this case 
\bd
  \tilde\omega^2(t)=\omega^2(t)-\frac{9H^2}{4},
\ed
which is time-dependent, thus one should expect particle production.} 
In this case we can then write the solution for $\chi_k$ as
\be
  \chi_k=\sqrt{\frac{1}{k}}e^{-ik|\eta|},
  \label{chi_conf}
\ee
which satisfies the initial conditions given in (\ref{a IC}). Note that (\ref{chi_conf}) coincides with 
the conformal vacuum. Using the definition of $\rho$, (\ref{drho_ex}), 
and (\ref{chi_conf}) we can see that 
\bd
  \rho=\frac{1}{\sqrt k}, \hs \rho_\eta=0. 
\ed
Finally using (\ref{N}) we see that the number of induced quasi-particles is thus
\bd
  N=0.
\ed

Using the usual Bogolyubov transformation method we can easily check that the so-called adiabatic 
vacuum (\ref{chi_0}) gives the same result as (\ref{chi_conf}). Thus, using (\ref{N_0}) we obtain
\bd
  N_{\chi_0}\sim\left|ik-ik\right|^2=0,
\ed
which is the same result. 

Finally, in the conformally coupled case, that is when $A=0$ and $y=1/2$, (\ref{R1}) gives the 
usual result
\bd
  \dot{\cal F}=\frac{\Delta E}{2\pi}\frac{1}{e^{2\pi\Delta E/H}-1},
\ed
which does have the expected $4D$ spectrum. Hence the response of the particle detector is 
that of an observer being in a thermal bath of radiation, even though, from above, there are 
no particles created. This is a well known result of the nonzero response of a comoving 
detector to a conformal vacuum \cite{Birrell:1982ix}.

\end{document}